# Mechanisms of rogue wave formation

Arnaud Mussot, Alexandre Kudlinski, Mikhail I. Kolobov, Eric Louvergneaux, Marc Douay and Majid Taki

*Université de Lille 1, Laboratoire PhLAM, IRCICA,*

*59655 Villeneuve d'Ascq Cedex, France*

*Corresponding author: mussot@phlam.univ-lille1.fr*

**Freak waves, or rogue waves, are one of the fascinating manifestations of the strength of nature. These devastating "walls of water" appear from nowhere, are short-lived and extremely rare. Despite the large amount of research activities on this subject, neither the minimum ingredients required for their generation nor the mechanisms explaining their formation have been given. Today, it is possible to reproduce such kind of waves in optical fibre systems. In this context, we demonstrate theoretically and numerically that *convective instability* is the basic ingredient for the formation of rogue waves. This explains why rogues waves are extremely sensitive to noisy environments.**

As told by seafarers, rogue waves are devastating "walls of water" appearing from nowhere [1,2]. They are short-lived and are often preceded or followed by deep holes in the surface of the sea. The few surviving observers also talk about "the three sisters", since they are usually composed of three successive waves. The probability of experiencing one of these terrible waves is very small but often leads to such a disaster that few eyewitnesses survive to describe them. Recently, scientific measurements have confirmed the stories of the survivors. Nowadays, there is a rapidly growing research activity devoted to improving models to successfully reproduce the oceanic observations of freak waves [1,2,3]. However, a full understanding of this phenomenon is far from having been achieved. Very recently, Solli *et al.* [4] have made considerable progress by observing an optical counterpart of oceanic freak waves in a photonic crystal fiber, producing a phenomenon coined the *optical rogue wave* (ORW) because of its similarity to the oceanic rogue wave. The authors of Refs. [4,5,6] have shown that the ORWs appear as solutions of the generalized nonlinear Schrödinger equation (GNLSE, see method) which is often used to describe the nonlinear dynamics of water waves as well as light propagation in photonic crystal fibers. This optical analogue could help a great deal in understanding the complex nonlinear behaviour of the rogue waves in the ocean. Indeed, the recent development of new technologies for the fabrication of microstructured optical fibers now allows us a great degree of freedom in the choice of the fiber parameters together with a high precision in their control.

Heretofore, experimental and numerical studies of ORWs have pointed out the crucial role of pulse propagation and the extreme sensitivity to initial noisy conditions, but no explanation of the mechanisms involving these ingredients has been given. We will demonstrate here that ORWs have their origin in *convective instabilities*, a type of instability that takes into account the drift effects in the process of wave amplification, resulting in extreme sensitivity to the input conditions. Furthermore, we demonstrate experimentally and numerically that ORWs can be generated by continuous-wave (CW) pumping conditions. This major difference with previous studies allows us to provide evidence for the formation of ORWs under conditions close to those for their hydraulic counterparts (calm, open ocean). We highlight the characteristics of these giant optical pulses (their brevity, their sudden appearance, their extremely high amplitude) as compared to common waves and discuss their similarity with freak waves [1,2,3]. Interestingly, one of the most striking results is that ORWs are surrounded by two satellites which look like the so-called "three sisters" in the hydraulic domain.

It is commonly accepted that fiber systems can be described by the GNLSE, making them very good candidates for parallels with water waves, also described by this equation. We have identified the terms in this equation which are physically responsible for the appearance of ORWs. In this minimal and relevant version, only the third-order dispersion ($\beta_3$) and the stimulated Raman scattering (SRS) effects are taken into account in the GNLSE. To emphasize their crucial role, we first performed numerical integrations of the standard nonlinear Schrödinger equation, *i.e.* with no SRS and no $\beta_3$. In order to obtain an insight into the statistical properties of the generated ORWs, we performed 100 numerical simulations corresponding to 100 different initial noisy conditions. The results are shown in Fig. 1(a-c). Figure 1(a) superposes the output spectra of each individual simulation (dotted grey lines) and the calculated mean spectrum (solid black line). We observed that the shape of the output spectra depends on the initial conditions, as is normal for a conservative nonlinear system. Relatively low-amplitude events (in comparison with common ORWs) occur over time [Fig. 1(b)] ,much as in the ocean [3]. However, the statistical distribution of these events does not have an "L" shape at all [here, an asymmetric bell-shaped curve in Log units, Fig. 1(c)], indicating that they cannot be interpreted as ORWs. Addition of the third-order dispersion and the stimulated Raman scattering terms in the previous equation drastically changes the dynamics of the system and rare high-amplitude events are then observed over time [Fig. 1(e)]. The peak power of these events is roughly up to 3-4 times larger than those of most of the other ones and, more importantly, their statistical distribution shows the characteristic "L-shaped" signature of the ORWs [4] [here an almost "L-shaped" curve even in Log units, Fig. 1(f)]. This amplitude ratio and the statistical features are clear signatures of ORWs, as reported in all previous studies. Thus, the third-order dispersion $\beta_3$ and the SRS are the basic ingredients for generating the ORWs.

Let us now describe the method employed to identify the nature of the instabilities at the origin of ORWs. Almost all recent observations pointed out that ORWs are unusually narrow and appear as sharp pulses recalling the "walls of water" in the ocean, with an extreme sensitivity to the initial conditions. This indicates, in particular, that ORWs possess a broad spectrum and hence cannot be described in the framework of classical linear stability analysis based on the normal-mode theory (plane waves). We have decided, therefore, that the stability analysis of ORWs has to be reformulated as an initial-value problem. The main advantage of this approach is that it allows us to understand the response of the system in relation to all types of localized or extended perturbations. More importantly, transport effects due to the propagation of an arbitrary perturbation at the input of the fiber are taken into account in the amplification process. As known from the literature [7,8], two different regimes of instabilities have been identified: *absolute and convective*. The drift velocity of the wave packet is usually used to distinguish both regimes. We calculated it from the GNLSE and we found that $V_{DRIFT}=f(\beta_3, SRS)$, see methods for details of the calculations. This means that when $\beta_3$ and Raman effects are present (GNLSE, Figs. 1-(d)-(f)) the instabilities are convective, while when their impact is negligible the instabilities are absolute (NLSE, Figs. 1-(a)-(c)) [9].

We are now going to describe the impact of the nature of the instability on the dynamics of the system. In the case of absolute instabilities, the initial perturbation at the system input becomes locally unstable at each spatial point of the system and in the laboratory frame. Thus the gain is more important than the drift, so the system becomes rapidly independent of the initial conditions. Therefore, the macroscopic output solutions are mainly determined by the dynamic nonlinear properties. In the case of convective instabilities, the initial perturbation at the system input does not grow locally but becomes unstable in a reference frame moving with the drift velocity. Thus the initial perturbation drifts away from its original position and finally leaves the system. This leads to a fundamentally different amplification mechanism where noise affects both the linear amplifications and the subsequent nonlinear dynamics. In this case, the random small localized perturbations are amplified and drift away while the new perturbations caused by the input noise are constantly seeded into the system and amplified in turn. The drift effect strongly increases the range of the initial conditions seeded into the system, particularly those rare random events which are responsible for rogue wave generation. On the contrary, these rare initial events are unlikely to be observed in the case of absolute instability where drift effect is absent and perturbations are amplified *in situ*. Thus, convective instabilities are responsible for the choice of the initial random events which will be amplified by the subsequent nonlinear dynamics to become macroscopic optical rogue waves.

Having demonstrated that the convective instabilities are at the origin of ORWs, we now proceed to describe their main characteristics (*i.e.* very high peak power and short duration) when, in a second step, nonlinear effects become dominant. In order to follow the convective nature of the ORWs where drift effects are crucial, we focus on the spectral and temporal dynamics of the wave propagation along the fiber (Fig. 2). Supercontinuum generation is nowadays one of the most appropriate optical configurations for observing ORWs, and the mechanisms of its dynamic evolution are relatively well known from the numerous studies performed in this field [11]. At the beginning of the fiber (approximately the first 100 m), the convective modulation instability process described above converts the CW field into a train of nonlinear pulses (hereafter, also called solitons) with comparable but not identical peak powers. The most powerful pulses [white circles in Fig. 2(a, b)] undergo the most efficient Raman self-frequency shifts, increasing the dispersion of the pulses and consequently reducing their velocities [12]. Due to the difference in the velocities for pulses of different power levels, collisions occur among these pulses [intersection between pulse curves in Fig. 2(b)]. As shown in Refs. [13,14,15], during a collision the most powerful pulse catches the energy from its less powerful neighbour. The amount of energy exchanged depends on the properties of the two initial pulses. A direct consequence of such a collision is a further enhancement of the Raman frequency shift [16] for the most powerful pulse. An example is shown in Fig. 2(a) where the most powerful pulses are accelerated and ejected from the wide central spectrum after about 200 - 300 meters of propagation. Once this strong nonlinear regime has been attained, the possible occurrence of a rogue wave event, as well as its spectral location, depends on the length at which the observation is performed [Fig. 2 (c)]. A close-up of one such event is represented in the inset in Fig. 2 (c). One can see that the peak power rapidly increases and decreases in less than a meter. The maximum peak power reached during the collision is 3 times higher than the mean peak power, fulfilling the general criterion for a rogue wave event. Focusing on the event appearing at a fiber length of 300 m, one can observe and provide evidence for the mechanism of an ORW formation by observing the temporal evolution of the spectrograms (Fig. 3). Two strong solitons (red patterns in Fig. 3), initially well-separated in the time domain [Fig. 3(a)], travel at different velocities until a collision occurs [Fig. 3(c)]. After colliding, they again behave independently [Fig. 3(e)]. When two solitons collide [Fig. 3(c)], a nonlinear interaction takes place, leading to the formation of a powerful spike of nearly 1 kW [central pulse in Fig. 3(f)]. This cluster spike is 3 times higher than the mean peak power and appears/disappears very quickly (in less than 1 meter, as seen in Fig. 3). This is the reason why we call this powerful pulse an ORW. Consequently, the mechanism of ORW formation can be identified as the collision between two already powerful solitons propagating at different velocities, as previously suggested in [17]. A very striking feature is the presence of two satellites around the central peak, recalling the famous "three sisters" (three consecutive freak waves) often mentioned by seafarers. This type of structure has not been reported in the pulsed regime because the most powerful soliton is rapidly ejected from the pulse and thus cannot collide with its neighbours. This fact strengthens the desirability of using CW or quasi-CW pumping in optical fibers in order to augment the correspondence between optical rogue waves and those observed in the ocean.

Our theoretical analysis is confirmed by the experimental results. Our experiment consists of launching the CW field produced by an Ytterbium fiber laser into a photonic crystal fiber with the same parameters as those used in the numerical simulations. The signal output spectra are recorded by a standard optical spectrum analyser with an integration time of the order of 1s. This means that the experimental spectrum represented in Fig. 4(a) corresponds to an averaging of millions of events. Its span from 850 nm to 1300 nm and its overall shape are similar to the numerical results in Fig. 1(d). In order to measure the probability of ORW generation, a spectral filter [transfer function represented in red in Fig. 4 (a)] is applied at the output end of the fiber, just in front of a high-bandwidth photodetector. This spectral filtering allows for the selection of highly nonlinear waves (the ones experiencing the most efficient red shift) and prevents the saturation of the photodetector by linear waves and pump residue. The single-shot time traces recorded with a high-bandwidth oscilloscope (25 GHz) are represented in Fig. 4(b). Rare and strong events clearly emerge from the background of pulses. The statistical distribution of these events as a function of their power is presented in Fig. 4(c). One can clearly observe a typical "L-shaped" curve attributable to the ORWs.

In conclusion, we have described the formation of optical rogue waves in a microstructured optical fiber under a continuous-wave pump regime. These initial conditions correspond to those of a calm ocean in which hydraulic freak waves are generated. Numerical simulations were performed to obtain a better understanding of the experiment and of the birth of optical rogue waves. The comparison between the experimental results and the numerical simulations has highlighted the physical mechanisms leading to the observation of optical rogue waves in this context. We have used a simple analytical model to demonstrate that they originate from convective instabilities, which explain their inherent extreme sensitivity to the initial conditions. The convective character of the nonlinear system is due to the drift effects (third-order dispersion and the spontaneous Raman scattering) accompanying the propagation of light in optical fibers. Additional numerical studies indicate that collisions between soliton-like pulses propagating at different velocities can induce extremely high-powered and sharp pulses, or optical rogue waves.

Our experiments with the CW laser together with the supporting numerical simulations reveal that the ORWs exhibit very specific features resembling those of ocean freak waves, such as the suddenness of their appearance and subsequent fading [1] and the presence of three consecutives high-powered pulses recalling the so-called "three sisters" in the ocean [1].

**Methods**

A simplified form of the GNLSE that includes only physical terms responsible of the ORW generation and dynamics is :

$$\frac{\partial E}{\partial z} = -i\frac{\beta_2}{2}\frac{\partial^2 E}{\partial \tau^2} + \frac{\beta_3}{6}\frac{\partial^3 E}{\partial \tau^3} + i\gamma E(z,\tau)\int_{-\infty}^{+\infty} R(\tau')|E(z,\tau-\tau')|^2 d\tau' - \alpha\frac{E}{2} \qquad (1)$$

Where $E(z,\tau)$ is the electric field envelope in a retarded time frame $\tau = t - \beta_1 z$ moving at the group velocity $1/\beta_1$ of the pump, $\beta_2$ and $\beta_3$ are the second- and the third-order dispersion terms respectively, and the last term with $\alpha$ describes the losses. $R(\tau)=(1-f_R)\delta(\tau) + f_R \mathrm{x} h_R(\tau)$ is the nonlinear response of silica ($f_R=0.18$ [12]). This expression is valid if we assume that the pulses are larger than a few tens of femtosecond in order to neglect the self-steepening effect. In the case of a fiber with a single-zero dispersion wavelength, when the spectral width of the field is not too broad, one can neglect the impact of the higher-dispersion terms. So that in addition to purely NLSE only the third-order dispersion and the Raman Effects are necessary for ORW occurrence. These two terms have a significant meaning that goes beyond the specific case of the NLS equation: they break the reflexion symmetry ($\tau \to -\tau$) giving rise to a systematic asymmetry in the solutions in both linear (third-order dispersion) and nonlinear (Raman effect) regimes. As a consequence, drift effects and high sensitivity to noise arise in the system. Equation 1 has been numerically integrated by using the adaptative step size method outlined by Sinkin et al. [18] with a local goal error set to $\delta_G=10^{-5}$. Initial quantum fluctuations are taken into account by adding a half-photon per mode at the fiber input. As demonstrated in Ref. [19] this leads to an accurate modelling of experimental noise in the generation of ORW's even in the case of an input CW pump wave. We used experimentally measured Raman response of silica fibers for $h_R(\tau)$. The spectral window was divided with $2^{15}$ points leading to a spectral resolution of 9.15 GHz, a time window of 0.109 ns and a temporal resolution of 3.3 fs. Note that we checked that neglecting the self-steepening effect as well as limiting the dispersion order up to 3 has no significant influence on the results.

For analytical predictions, the classical linear stability theory is insufficient, as it stands, to explain the ORW generation since it applies to extended perturbations characterized by a single wavenumber. By contrast, in order to determine the linear response of the system to a localized perturbation, it is necessary to include a finite band of modes in the dynamical description. This can be achieved by reformulating the linear stability analysis as an initial-value problem. For a detailed description of the concept and techniques of instabilities in terms of an initial-value problem analysis (including absolute and convective instabilities), the reader is referred to the original work by Briggs [20] and the recent works in optics [21]. Starting from Eq. (1), the response of the system to any perturbation around the tau-independent CW nonlinear solution is obtained via the usual Fourier integral over the whole range of frequencies. The carrier frequency, wave number, and velocity of the most convectively unstable wave packet are determined following the method of steepest descend [22]. The latter involved rather complicated algebraic formulas and will be published in more specialized review. To give more insight about the propagation effect resulting from convective instability we present a simplified expression of the wave packed velocity when only third order dispersion is taken into account (no Raman Effect). It reads:

$$1/V_{DRIFT} = -I_0(\beta_3/|\beta_2|) \pm \sqrt{2[I_0(\beta_3/|\beta_2|)]^2 + 4|\beta_2|I_0}$$

where $I_0$ is the CW input power. Note the symmetry breaking in the above expression introduced by third-order dispersion term.

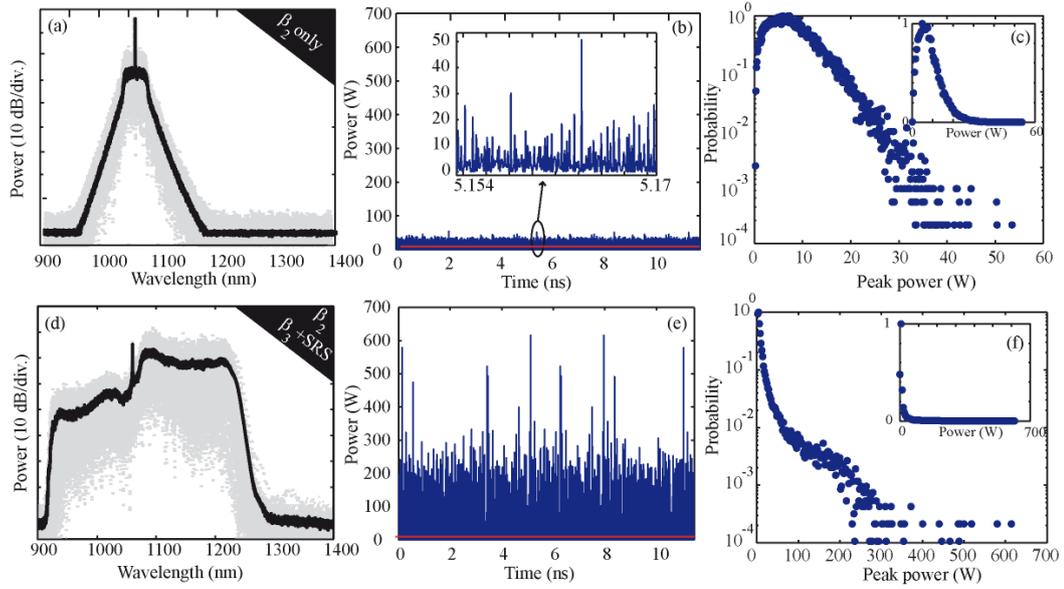

Figure 1 : Collection of 100 numerical simulations resulting from different initial noise conditions (a) to (c) with only $\beta_2$, (d) to (f) with $\beta_2$, $\beta_3$ and Raman terms. (a) and (d) numerical spectra (grey dotted lines) with the averaging in solid black curve. (b) and (e) the corresponding 100 simulations in the time domain arranged end to end. (c) and (f) associated histogram with 100 bins (c) and 2 W (f). Parameters of the simulation : P=10 W, $\beta_2=-2.6.10^{-28}$ s²/m, $\beta_3= -0,7.10^{-40}$ s³/m, $\gamma=11$ W$^{-1}$.km$^{-1}$, $\alpha=10$ dB/km and L=400 m.

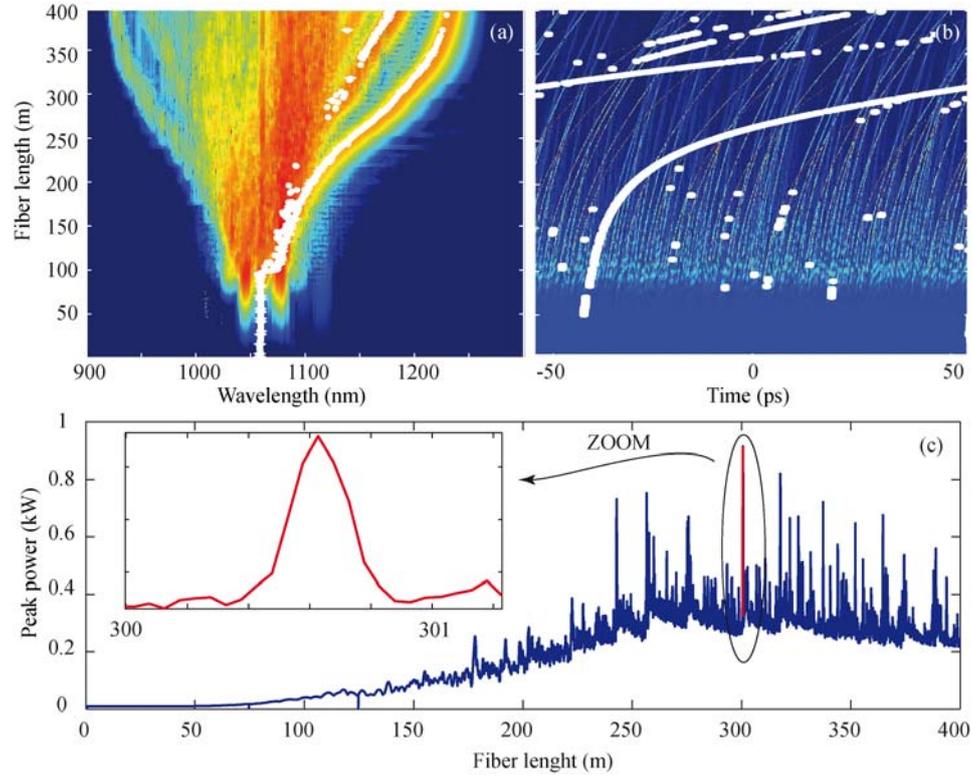

Figure 2: Longitudinal spectral (a) and time (b) evolution of one of the events represented in Fig. 1 (lower subfigures). The white dots represent the most powerful solitonic pulse. (c) Evolution of the peak power over the fiber length.

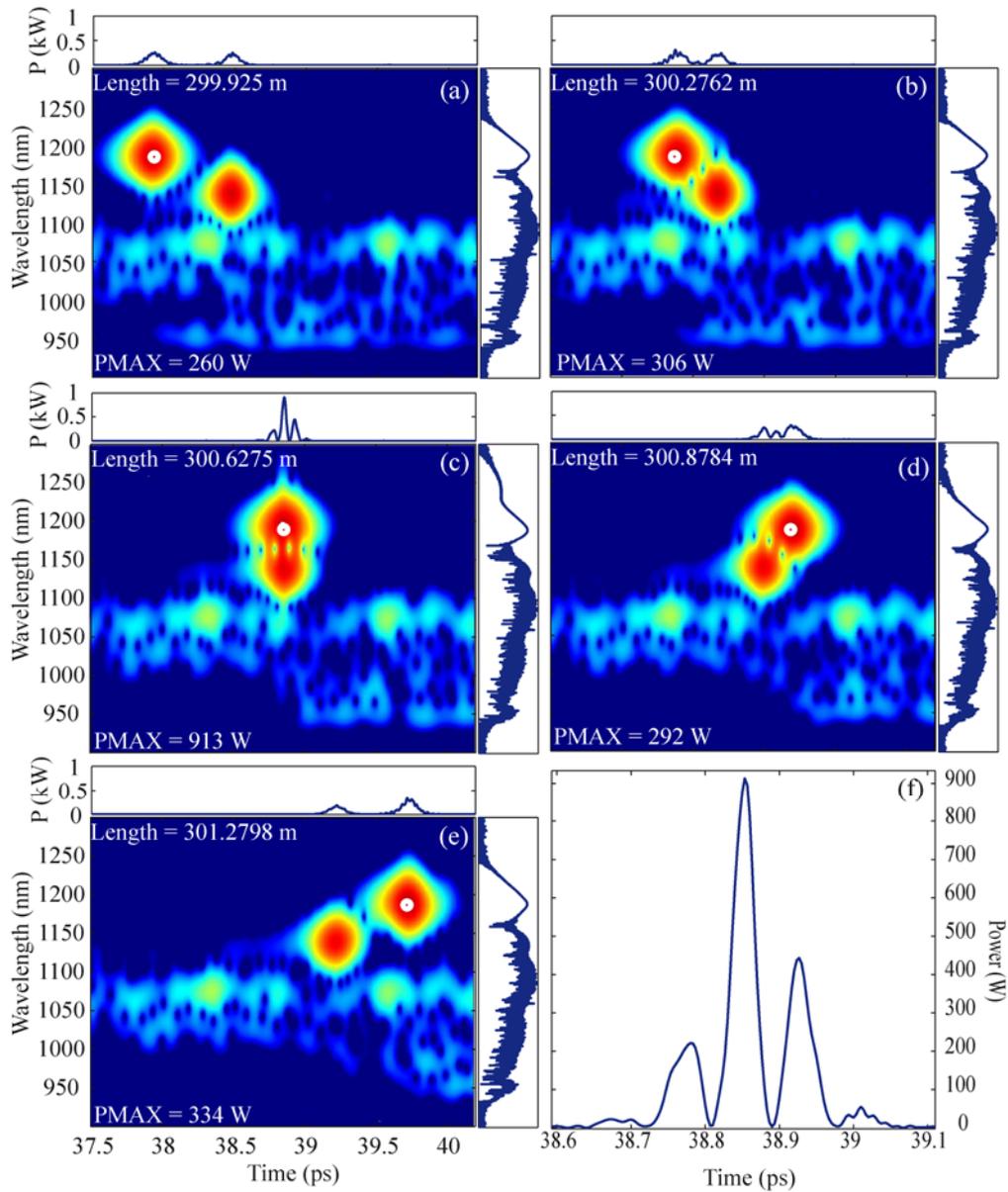

Figure 3: Spectrogram illustrating the appearance/disappearance of a rogue wave from a fiber length L=300 m to L=301 m.

17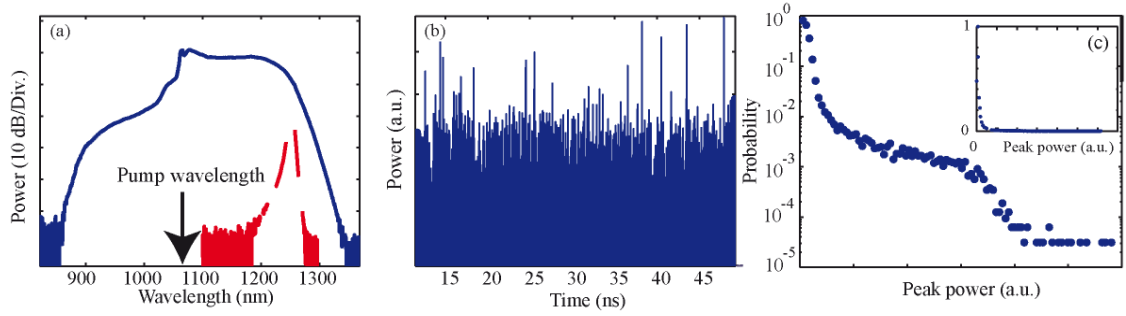

Figure 4 : Experimental results. (a) Output spectrum. (b) Single-shot time trace and (c) associated histogram.

**Supplementary Information**

The movie intituled *spectrogramORW* shows a spectrogram representing the formation of ORGs from the beginning of the fiber untill the end (QuickTime; 9.6 MB).

The movie intituled *spectrogramORWzoom* shows a spectrogram representing a zoom (from L=300 m to L=303 m) on the collision between two soliton leading to the formation of an ORG (QuickTime; 6.6 MB).